\documentclass[twocolumn]{aastex631}

\newcommand\lsim{\mathrel{\rlap{\lower4pt\hbox{\hskip1pt$\sim$}}
\raise1pt\hbox{$<$}}}
\usepackage{xcolor}

\shorttitle{Stellar Bumper Cars}
\shortauthors{Rose \& Mockler 2024}

\graphicspath{{./}{figures/}}

\begin{document}


\title{On the Orbital Effects of Stellar Collisions in Galactic Nuclei: \\
Tidal Disruption Events and Ejected Stars} 

\correspondingauthor{Sanaea C. Rose}
\email{sanaea.rose@northwestern.edu}

\author[0000-0003-0984-4456]{Sanaea C. Rose}
\affiliation{Center for Interdisciplinary Exploration and Research in Astrophysics (CIERA), Northwestern University, 1800 Sherman Ave, Evanston, IL 60201, USA}

\author[0000-0001-6350-8168]{Brenna Mockler}
\affiliation{The Observatories of the Carnegie Institution for Science, Pasadena, CA 91101, USA}

\begin{abstract}

Dense stellar clusters surround the supermassive black holes (SMBH) in galactic nuclei.
Interactions within the cluster can alter the stellar orbits, occasionally driving a star into the SMBH's tidal radius where it becomes ruptured, or expelling a star from the nuclear cluster. This proof-of-concept study examines the orbital effects of stellar collisions using a semi-analytic model. Both low and high speed collisions occur in the SMBH's sphere of influence.  
We find that collisions can place stars on nearly radial orbits. Depositing stars within the tidal radius, collisions may drive the disruption of stars with unusual masses and structures: depending on the nature of the collision, the star could be the product of a recent merger, or it could have lost its outer layers in a previous high speed impact, appearing as a stripped star.
We also find that high speed collisions near the periapsis of an eccentric orbit can unbind stars from the SMBH. However, dissipation during these high-speed collisions can substantially reduce the number of unbound stars achieved in our simulations. We conclude that tidal disruption events (TDEs) and ejected stars, even in the hypervelocity regime, are plausible outcomes of stellar collisions, though their frequency in a three-dimensional nuclear star cluster are uncertain. Future work will address the rates and properties of these events.




\end{abstract}



\section{Introduction}

A supermassive black hole resides at the center of most galaxies, where it is surrounded by a dense stellar cluster \citep[e.g.][]{FerrareseFord05,KormendyHo13,schodel+03,Ghez+05,Ghez+08,Gillessen+09,Gillessen+17,Neumayer+20}. A tidal disruption event (TDE) occurs when a star from the cluster passes within a critical distance from the SMBH and becomes ruptured by tidal forces \citep[e.g.,][]{Hills1975,Rees1988,Alexander99,MagorrianTremaine99,WangMerritt04,macleod_tidal_2012}. The SMBH then accretes the stellar material, producing an electromagnetic signature \citep[e.g.,][]{Guillochon+13}. Spectra of these events encode valuable information about the mass, structure, and composition of the ruptured star \citep[e.g.,][]{kochanek_abundance_2016, kochanek_tidal_2016, Yang17,Mockler+22,Miller+23}. Observations of TDEs represent a powerful way to probe the stellar populations in galactic nuclei and the processes that shape them.


Amongst these physical processes are direct collisions, which occur within the sphere of influence of the SMBH due to the high densities and velocity dispersion \citep[e.g.,][]{FreitagBenz02,Dale+09, DaleDavies,RubinLoeb,Balberg+13,Balberg24,Mastrobuono-Battisti+14,RoseMacLeod24}. These events may produce electromagnetic signatures, both from the collisions themselves and from interactions between liberated material and the SMBH \citep[e.g.,][]{rosswog_tidal_2009, lee_short_2010,Balberg+13,Dessart+24,Ryu+24a,Ryu+24b,Brutman+24}, though it is difficult to ignite a main-sequence star with compression \citep{Guillochon+09}. They can also shape the stellar population by altering the masses of the stars and in some cases giving rise to blue stragglers \citep[e.g.,][]{Lai+93,Rauch99,Sills+97,Sills+01,Lombardi+02,macleod_spoon-feeding_2013, Leigh+16,Rose+23}. Recently, \citet{Gibson+24} have shown that very high-speed collisions can also produce stripped stars similar to what might be seen through binary evolution. 

Collisions that produce stellar mergers or stripped stars may explain recent unexpected TDE observations.
For example, detections of high nitrogen-to-carbon (N/C) ratios in TDEs point to the disruption of more stars that burn on the CNO cycle \citep[as first proposed by][]{kochanek_abundance_2016, kochanek_tidal_2016}, and that are $\gtrsim 1-2 M_\odot$ \citep{kochanek_abundance_2016, Yang17,gallegos-garcia_tidal_2018, Mockler+23} than is predicted by the host galaxies' stellar populations \citep{Mockler+23}.   
One particular TDE has such an extremely high N/C abundance ratio that it is difficult to explain with single stellar evolution alone \citep[][]{Miller+23}, but could be the result of the disruption of a stripped star that has lost its nitrogen-poor envelope \citep[][]{Mockler+24}. 


The potential link between TDEs and collisions is intriguing: collisions can affect both the properties of the stars and their orbits about the SMBH. \citet{macleod_tidal_2012} first considered collisions, particularly destructive ones, in the context of the TDE rate.
Changes to a star's trajectory from collisions have also been considered in other dynamical models of dense stellar systems \citep[e.g.,][]{Sanders1970,Rauch99,FreitagBenz02,Kremer+20,Gonzalez+21,Rodriguez+22}, but remain under-explored in the context of nuclear star clusters. 

Beyond the potential connection to TDEs, collisions can have significant implications for the stellar orbits. Within $0.1$~pc of the Milky Way's SMBH, collisions act on a shorter characteristic timescale compared to two-body relaxation, and collisions have long been recognized as an important factor in shaping the stellar cusp \citep[e.g.,][]{DuncanShapiro83,Murphy+91,David+87a,David+87b,Rauch99,BalbergYassur23,RoseMacLeod24,AshkenazyBalberg24}. Furthermore, as mentioned above, collisions can produce unusual stars, either stripped in a high-speed collision or the product of a merger, and the final orbital properties can indicate where these stars might be observed in the Milky Way's Galactic center.
In this proof-of-concept study, we study the orbital effects of stellar collisions in galactic nuclei. We assess whether the collisions that produce unusual stars could plausibly deposit those same stars onto TDE-producing orbits. Our models leverage simple, intuitive treatments for collision outcomes and fitting formulae from previous studies \citep[e.g.][]{Lai+93,Rauch99}. We test a range of initial conditions and nuclear star cluster properties. Our paper is organized as follows:

In Section~\ref{sec:modelcluster}, we discuss our general approach to modeling the Milky Way's nuclear star cluster. Section~\ref{sec:ICs} describes the treatment of various physics in our code, with Section~\ref{sec:collisionorbits} in particular outlining our methodology for updating the stellar orbits post collision.
Section~\ref{sec:results} presents simulated results for direct collisions. Sections~\ref{sec:TDEs}, \ref{sec:orbitalshaping}, and \ref{sec:unbound} discuss the implications for TDEs, orbital properties of collision-affected stars, and unbound and hypervelocity stars. We then incorporate relaxation into our simulations in addition to stellar collisions and present results in Section~\ref{sec:withrelaxation}. Lastly, we summarize the scope and findings of our study in Section~\ref{sec:conclusions}.



\section{Model Nuclear Star Cluster} \label{sec:modelcluster}

We leverage semi-analytic models to study the effects of collisions on the nuclear star cluster. Our fiducial model uses the conditions and properties of the Milky Way's Galactic nucleus (GN), whose proximity makes it the best studied of these environments. We follow a sample of stars embedded in a fixed, unevolving cluster, detailed in this section. For simplicity, both the evolving sample and the surrounding cluster are composed of $1$~M$_\odot$ stars. The cluster can be understood in two key properties, density and velocity dispersion, which govern the dynamical processes unfolding within it.

The stellar density sets the the frequency with which stars interact. We describe the stellar density as a function of distance from the SMBH using a power law:
\begin{eqnarray} \label{eq:density}
    \rho(r_\bullet) = \rho_0 \left( \frac{r_\bullet}{r_0}\right)^{-\alpha} \ , 
\end{eqnarray}
where $\alpha$ is the slope and $r_\bullet$, distance from the SMBH. Based on observations of the cluster within the sphere of influence, this equation is normalized using $\rho_0 = 1.35 \times 10^6 \, M_\odot/{\rm pc}^3$ at $r_0 = 0.25 \, {\rm pc}$ \citep{Genzel+10}. Our fiducial model uses a slope of $1.75$, the expectation for a single-mass population \citep{BahcallWolf76}, consistent with the fact that our simple model cluster has only solar mass stars, though we note that with a range of masses and populations, our Galactic center may have a shallower stellar cusp \citep[e.g.,][]{Gallego-Cano+18,LinialSari22}. 
We assume that the slope of the stellar cusp is roughly constant, or varying slowly, over the timescales of interest in our simulations.

The velocity dispersion within the cluster also influences the frequency and nature of stellar interactions. It decreases with distance from the SMBH:
\begin{eqnarray}\label{eq:sigma}
    \sigma(r_\bullet) = \sqrt{ \frac{GM_{\bullet}}{r_\bullet(1+\alpha)}},
\end{eqnarray}
where $\alpha$ is the slope of the density profile and $M_{\bullet}$ is the mass of the SMBH \citep{Alexander99,AlexanderPfuhl14}. We take $M_\bullet$ to be $4 \times 10^6$~M$_\odot$, like the Milky Way's SMBH \citep[e.g.,][]{Ghez+03}. For a uniform mass cluster of $1$~M$_\odot$ stars, the number density $n$ is simply $\frac {\rho(r_\bullet)}{1 \, M_\odot}$.

\section{Semi-analytic Model} \label{sec:ICs}
We follow a sample of $1$~M$_\odot$ stars embedded in our model cluster. Our simulations without two-body relaxation follow a sample of 10,000 stars, while our more computationally intensive runs with two-body relaxation follow 4,000 stars.
The orbital properties of the sample stars are drawn so that their statistical distributions are consistent with the cluster population at large.
The orbital eccentricities have a thermal distribution, while we select their semimajor axes so that they lie on a cusp with slope $\alpha$, matching the background cluster. 
These stars are allowed to evolve under the influence of two main dynamical processes, direct collisions and two-body relaxation, using a model first developed by \citet{Rose+22,Rose+23}. The conditions of the surrounding stellar cluster, described in the previous section, determine the probability that a given star will experience a collision. 
Therefore, while we only follow a representative sample of stars, their evolution accounts for the presence of 4 million stars in the inner parsec.
Furthermore, because our sample’s spatial distribution is the same as that of the cluster, it allows for easy scaling of the results, as described at the beginning of Section~\ref{sec:results}. We account for the following physical processes:

\subsection{Collisions} \label{sec:collisionsgeneral}
\begin{figure}
	\includegraphics[width=0.95\columnwidth]{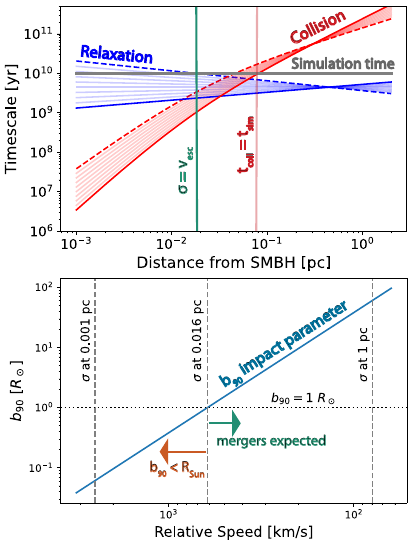}
	\caption{\textbf{Upper Panel:} We plot the relevant timescales as a function of distance from the SMBH for a range of stellar density profiles, $\alpha = 1.25$ (solid line) to $\alpha = 1.75$ (dashed line). The collision and relaxation timescales are in red and green, respectively, while the grey line marks the total simulation time of $10$~Gyr. The vertical red line emphasizes the radius at which the collision timescale equals the simulation time; within this radius, collisions are common and shape the evolution of the stars and cluster. We also mark the radius at which the velocity dispersion is approximately the escape speed from the surface of a Sun-like star using the green vertical line. External to this radius, we expect collisions to result in mergers. \textbf{Lower Panel:} We consider a hyperbolic encounter between two stars in the cluster. This plot shows the impact parameter in solar radii for a $90 ^{\circ}$ deflection as a function of the relative speed between the stars. Note that the x-axis is inverted to reflect the dependence of velocity dispersion on distance from the SMBH; it parallels the x-axis of the plot above, with the largest velocities occurring near the SMBH and decreasing further out. We have marked the speeds that correspond to the inner and outer allowed initial semimajor axes of the stars in our sample, $0.001$ and $1$~pc, respectively. Additionally, we mark where $b_{90}$ becomes equal to the star's radius. This point coincides well with the speed at which we would expect the collision outcome to transition from mergers to the sorts of high-speed encounters that might strip or destroy stars, but leave them unbound from each other.}
     \label{fig:timescales}
\end{figure}

Direct collisions occur over a characteristic timescale $t_\mathrm{coll}^{-1} = n \sigma A$, where $A$ is the cross-section of interaction, $n$ is the number density, and $\sigma$ is the velocity dispersion. 
Specifically, $A$ is the physical cross-section enhanced by gravitational focusing. The collision timescale also depends weakly on the star's orbital eccentricity 
and can be written as:
\begin{eqnarray} \label{eq:t_coll_main_ecc}
     t_{\rm coll}^{-1} &=& \pi n(a_\bullet) \sigma(a_\bullet) \nonumber \\ &\times& \left(f_1(e_\bullet)r_c^2 + f_2(e_\bullet)r_c \frac{2G(M_\odot+M_{star})}{\sigma(a_\bullet)^2}\right)\ .
\end{eqnarray}
where $f_1(e_\bullet)$ and $f_2(e_\bullet)$ are equations 20 and 21 from \citet{Rose+20}, $G$ is the gravitational constant, and $a_\bullet$ is the star's semimajor axis. $M_{star}$ represents the mass of the star which we are following, while $M_\odot$ is the mass of a star from the cluster with which it collides. Recall that we assume the cluster is composed entirely of $1$~$M_\odot$ stars. At the beginning of our simulation, $M_{star} = M_\odot$ as well, though its mass can change if it experiences collisions during the simulation.  The parameter $r_c$ is the sum of the radii of the colliding stars, or $2R_\odot$ for a uniform population of solar mass stars. We plot this timescale in red in the upper panel of Figure~\ref{fig:timescales} for a range of slopes for the stellar density profile, spanning $\alpha = 1.25$ (dashed line) to $\alpha = 1.75$ (solid line). The horizontal grey line shows the total simulation time, included to guide the eye. Where the collision timescale is less than the simulation time, within $0.1$ pc of the SMBH, collisions become important to understanding the evolution of the cluster \citep[e.g.,][]{RoseMacLeod24}.

We treat stellar collisions using a statistical approach. We begin by computing the probability that a star in our sample will experience a collision. Over a timestep $\Delta t$, this probability equals $\Delta t/t_{coll}$. Over the parameter space we consider, $t_{coll} \gtrsim 10^7$ years, and so we set $\Delta t = 10^6$ to ensure $t/t_{coll}$ is always less than one. 
The code then draws a random number between $0$ and $1$, which, if less than or equal to the collision probability $\Delta t/t_{coll}$, means a collision has occurred.
We repeat this prescription until the desired simulation runtime or the star's main-sequence lifetime has been reached, whichever is shorter. Stars can and do experience multiple collisions over their lifetime in our simulations, in particular if $t_{coll}$ is less than the lifetime of the star.

If a collision has occurred over a given timestep, we update the mass and age of the star using a prescription detailed in \citet{Rose+23}. A full discussion of our approach can be found in previous papers \citep{Rose+23,RoseMacLeod24}. In brief, we combine fitting formulae from hydrodynamics simulations of stellar collisions \citep{Rauch99,Lai+93} with heuristic arguments to (1) determine if a given collision will result in a merger and (2) determine the amount of mass lost from either the merger product or the unbound individual stars. 
Mergers with minimal mass loss are most likely to occur outside of about $0.01$~pc, where the velocity dispersion is less than the escape speed from the stars. Within $0.01$~pc, however, velocities exceed the escape speed from the stars, and collisions can result in peculiar, low-mass ``stripped stars'' \citep[][]{Rose+23,RoseMacLeod24,Gibson+24}.

In practice, the precise outcome of a collision is more complex to determine, depending on properties such as the impact parameter and stellar structure \citep[e.g.,][Rose et al. in prep.]{FreitagBenz,Gibson+24}. However, the results of these smooth particle hydrodynamics (SPH) studies follow the trends described above \citep[see full comparison in][]{Rose+23}. In this study, which focuses on the general implications of collisions for stellar orbits, we test two different SPH-based prescriptions to determine the outcome of each collision. Simulations which use fitting formulae from \citet{Rauch99} for the mass loss and escape speed arguments to determine whether or not a collision results in a merger are labeled as ``Rauch99'', while those that use fitting formulae from \citet{Lai+93} are labeled as ``Lai+93''.

In previous iterations of this code, the collision speed was taken to be simply the velocity dispersion at the star's distance from the supermassive black hole, given by Eq.~(\ref{eq:sigma}) \citep{Rose+22,Rose+23}. However, this approach is insufficient for studying the effects of collisions on stellar orbits. In this study, we statistically draw both the collision location along our sample star's orbit and the velocity of the second colliding star from the background cluster. We note that we only use the word orbit to refer to orbits about the SMBH, not between the two colliding stars, unless otherwise specified.

\subsection{Orbital Dynamics} \label{sec:collisionorbits}

If a star in our sample with semimajor axis $a_\bullet$ and eccentricity $e_\bullet$ experiences a collision over our timestep $\Delta t$ at some point during the simulation, we draw the location along its orbit where the collision occurs. The probability that a collision will occur over some increment of time can always be estimated as $\delta t/t_{coll}$, where we use $\delta$ to distinguish this arbitrary interval from our simulation timestep $\Delta t$. Taking $\delta t \rightarrow dt = dr_\bullet/v_{r_\bullet}$, where $v_{r_\bullet}$ is the radial velocity, the probability density $p_{coll}$ as a function of radial distance $r_\bullet$ would be proportional to $t_{coll}^{-1} dr_\bullet/v_{r_\bullet}$. At apoapsis and periapsis both $v_{r_\bullet}$ and $dr_\bullet$ go to zero, but the integral can be regularized by transforming to the mean anomaly. $t_{coll}^{-1}$ is tantamount to a flux of colliders experienced by the sample star as it traverses its orbit. For $t_{coll}$, we tested both Eq.~\ref{eq:t_coll_main_ecc} as is and with $\sqrt{v^2+\sigma^2}$ in place of $\sigma$, where $v$ is the speed of the star in question and a function of $r_\bullet$. The results were almost indistinguishable, but we use the latter in our simulations. While a star spends most of its time near apoapsis -- the probability density of finding a star at some $r_\bullet$ along its orbit is proportional to just $dr_\bullet/v_{r_\bullet}$ -- \textbf{it is} actually more likely to collide near periapsis because of the steep, cuspy stellar density profile, consistent with, e.g., \citet{SariFragione19}. Given the probability of collision as a function of $r_\bullet$, we draw the collision location of the star using inverse transform sampling.

Once we have the collision radius, we then must draw a plausible velocity for the second colliding star. Both theoretical arguments and observations suggest that the nuclear star cluster is isotropic and Maxwellian
\citep[see the review][and citations therein]{Alexander05}. We draw the relative speed between our sample star and the collider assuming a Maxwellian velocity distribution with dispersion $\sigma$ \citep[e.g.,][equation 8.46]{BinneyTremaine}. We note that simply drawing $\mathbf{v_\mathrm{collider}}$ from a Maxwell-Boltzmann distribution with dispersion $\sigma$ yields qualitatively similar results, where $\sigma$ is from Eq.~\ref{eq:sigma} evaluated at the collision radius. We assume the velocity of the collider is isotropically distributed \citep[for a similar technique of applying isotropic velocity kicks, see][]{Jurado+24}. As our model cluster is composed of a uniform stellar population, the mass of the collider $m_{collider}$ is always $1$~M$_\odot$.

The relative speed tells us whether or not a merger occurs and the degree of fractional mass loss from the system (see above section). From here, determining the final orbit(s) will depend on the type of outcome. 

\subsubsection{Final Orbit in Merger Case} \label{sec:finalorbitmerger}

A stellar merger is the natural outcome of a low-speed collision. We calculate the final mass, age, and new trajectory of the product assuming the stars act as ``sticky spheres'' \citep[e.g.,][]{Rodriguez+22,Rose+23}. The two stars approach each other with velocity vectors $\mathbf{v_\bullet}$ and $\mathbf{v_{collider}}$.
Momentum conservation demands that the final velocity of the merger product, $\mathbf{v_{final}}$, equals $ \left( M_{star} \mathbf{v_\bullet} + M_\odot \mathbf{v_{collider}} \right)/(M_{star}+M_\odot)$. With the final velocity, mass, and collision position, we can calculate the new orbit of the merger product about the SMBH. As mergers only occur at lower speeds -- heuristically, the relative speed must be less than the escape speed from the star -- the mass loss from these collisions tends to be low, as indicated by hydrodynamics studies \citep[e.g..,][]{Lai+93,Rauch99,FreitagBenz}. However, we note that in cases of colliding stars with different masses, asymmetric mass loss can impart a velocity kick to the merger product. For this proof-of-concept, we assume that given the low mass loss from mergers, the kick will similarly be small. Previous hydrodynamics studies have found stellar merger kicks in globular clusters to be low, at most $\sim 10$ km/s \citep[e.g.,][]{Lombardi+96}, however our future work will fully assess the limits of the sticky sphere approach as applied to nuclear star clusters.

\subsubsection{Final Orbits Following a High-Speed Collision} \label{sec:HSorbitcalc}

Here we present a framework for the analytical treatment of orbital changes from high-speed collisions.
In order to understand these collisions, we start at the limit where the two stars barely graze. This interaction should unfold as a hyperbolic encounter. In the center of mass frame of the two stars, their speeds remain constant, but their velocity vectors are deflected by angle $\theta_{hyp}$. $\theta_{hyp}$ can be found analytically:
\begin{eqnarray} \label{eq:theta}
\theta_{hyp} = 2 \arctan(b_{90}/b) \ ,
\end{eqnarray}
where $b$ is the impact parameter and $b_{90}$ is defined as the impact parameter needed for a $90 ^{\circ}$ deflection. $b_{90}$ equals $ G(M_\odot+M_{star})/v_{rel}^2$, where $v_{rel}$ is the relative speed between the two stars \citep[e.g.,][equation 3.52]{BinneyTremaine}.

We plot $b_{90}$ as a function of the relative speed in the bottom panel of Figure~\ref{fig:timescales}. In the nuclear star cluster, the velocity dispersion can be understood as the characteristic relative speed between stars at a given distance from the SMBH (Eq.~\ref{eq:sigma}). We have therefore inverted the x-axis of the bottom plot to parallel that of the upper plot, distance from the SMBH, and marked the velocity dispersion at key distances with vertical dashed lines. In both the upper and lower plots, we also indicate the regions in which we expect mergers versus high-speed collisions, which leave the stars unbound from each other. Interestingly, for most of the parameter space where these high-speed collisions occur, $b_{90}$ is less than the star's radius. Another way of interpreting this statement is that at high speeds, physical collisions are required for a strong-angle deflection \citep[see also][]{Alexander05}.

We base our treatment of high speed collisions on hyperbolic encounters between stars. In galactic nuclei, gravitational focusing between stars is generally weak because of the high relative velocities. 
Therefore, the distance between the two stars upon physical impact is comparable to the impact parameter as the stars approach each other from infinity. From hereon, we use impact parameter $b$ to equivalently denote the distance between each star's center when they collide. Our $b$ is therefore defined as always less than the sum of the radii of the two stars, and in our simulations we draw $b$ statistically for each collision from a probability density $p(b) \propto \frac{b}{r_c^2} db$.

If the stars were point particles, they could get arbitrarily close and the interaction would still unfold as a hyperbolic encounter. Two Sun-like stars begin to touch when $b = 2$~R$_\odot$. With $b < 2$~R$_\odot$, the stars physically impede each other as they interact, leading to a smaller deflection angle than the one given in Eq.~(\ref{eq:theta}). Furthermore, if the stars approach each other perfectly head-on with $b = 0$, the center of mass velocity is $0$ and there is no angular momentum. In this case, there would be no deflection. Heuristically, then, we expect the deflection angle to be given by Eq.~(\ref{eq:theta}) for grazing encounters, and some fraction of this angle for encounters with $b< 2R_\odot$. That fraction should decrease with impact parameter until they are both zero. The precise dependence can be tuned by hydro simulations, but for this proof-of-concept study, we define a collision deflection angle that meets the two limiting criterion:
\begin{eqnarray} \label{eq:colltheta}
\theta_{coll} = 2 \frac{b}{r_c} \arctan(b_{90}/b) \ ,
\end{eqnarray}
where $r_c$ is the sum of the radii of the two colliding stars. This equation leads to more conservative deflection angles compared to Eq.~\ref{eq:theta} and suppresses deflections when the stars significantly overlap. 
It is reasonable to expect that the collision affects the stars' speeds as well. Even more so than the deflection angle, changes in speed can only be understood through hydrodynamic simulations. In our models, we simply test three cases: one in which the speed is not affected at all, one in which the speed in the center of mass frame is always reduced by $10 \%$, and one in which it is always reduced by a factor of $2$.

We treat these collisions as follows: we first calculate the center of mass velocity of the two stars. We transform to their center of mass frame. Only then do we rotate and scale their velocity vectors. 
We then transform back to the frame of the SMBH and calculate the new orbits given the velocity and position vectors. We note that our calculation assumes that the center of mass is not moving in the frame of the supermassive black hole. In actuality, the center of mass of the two stars would orbit the supermassive black hole, but the effects will be negligable due to the high collision speeds.

\subsection{Two-Body Relaxation}

In addition to collisions, stars in the cluster experience the weak gravitational effects of nearby neighbors. The effects of these interactions can accumulate, eventually changing the star's orbital energy and angular momentum by order of itself. The original orbit is ``erased'' over a characteristic timescale:
\begin{eqnarray} \label{eq:t_rlx}
t_{\rm rlx} = 0.34 \frac{\sigma^3}{G^2 \rho \langle M_\ast \rangle \ln \Lambda_{\rm rlx}} \ ,
\end{eqnarray}
where $\langle M_\ast \rangle$ is the average star's mass, here taken to be $1$~M$_\odot$, and $\ln \Lambda_{\rm rlx}$ is the coulomb logarithm \citep[e.g.,][]{BinneyTremaine,Merritt2013}. Figure~\ref{fig:timescales} shows the relaxation timescale as a function of distance from the SMBH in blue for a range of stellar density profiles. 


We account for relaxation by allowing the orbital eccentricity and semimajor axis of each of our sample stars to slowly evolve. Once per orbit, we apply a small instantaneous change in velocity to each star \citep[e.g.,][see the latter for the full set of equations]{Bradnick+17,Lu+19,Rose+22,Rose+23,Naoz+22}.
The kick is calibrated so that $\Delta v/v \sim \sqrt{\Delta t/t_{rlx}}$, and if $\Delta t = t_{rlx}$, $\Delta v$ is of order of the velocity. 
This prescription simulates the diffusion of the orbital orbital parameters over time from interactions with other stars in the cluster. Previously, it has been used in studies of TDEs and extreme mass ratio inspirals of stellar mass black holes into the SMBH \citep{Naoz+22,Melchor+24}.


\subsection{Orbital Stopping Conditions} \label{sec:stoppingconditions}

As noted in Section~\ref{sec:collisionsgeneral}, we terminate the simulation when the desired runtime of $10$ billion years is reached or when the time elapsed has exceeded the star's main-sequence lifetime, whichever comes first. However, orbital changes from collisions or relaxation can also send stars into the tidal radius, where they will be destroyed. We trigger a stopping condition if the star's periapsis $a_\bullet(1-e_\bullet)$ becomes less than twice the tidal radius from the SMBH, $R_{star} \left( M_\bullet/M_{star} \right)^{1/3}$ \citep[e.g.,][]{Guillochon+13,Mockler+23}. Tidal disruption events can be characterized by impact parameter $\beta = R_{tidal}/(a_\bullet(1-e_\bullet))$. Our stopping condition corresponds to $\beta = 0.5$, allowing us to capture partial as well as full disruptions.

Because both direct collisions and relaxation processes can place a star onto a nearly radial orbit, we log whether the critical orbit was reached through a direct collision or our relaxation prescription. 
A star that becomes a TDE due to relaxation processes can still have experienced one or more collisions previously in its life. However, a star that collides and becomes a TDE due to the collision itself may still be inflated from the impact when it reaches its periapsis. In this case, the stopping condition noted above may be conservative; the $R_{star}$ is simply the radius expected using a mass-radius relation for a main-sequence star of mass $m_{star}$. Our models use $R_{star} = 1.01R_\odot \times (M_{star}/M_\odot)^{0.57}$ from \citet{DemircanKahraman91}, which yields similar radii as their empirical fit for stars with mass between $0.1$ and $18$ M$_\odot$. We do not allow the stellar radii to evolve with the age of the star, and plan to explore a more holistic treatment of the stellar evolution in future work.

High-speed collisions, unlike those that result in mergers, can also cause stars to be ejected from the nuclear star cluster. Consider two stars that collide on elliptical orbits about the SMBH that intersect near periapsis. As the eccentricity approaches unity, the speed at periapsis approaches the escape speed, just shy of becoming an unbound, parabolic orbit. In the limit of a hyperbolic encounter, the speeds of the stars are unchanged in the center of mass frame, but they undergo a deflection. In the frame of the SMBH, this change can boost one star's speed enough to unbind it from the SMBH, while the other star ends up on a more tightly bound orbit. Close encounters have previously been shown to eject stars from dense stellar systems \citep[e.g.,][]{Henon69,LinTremaine80}. Additionally, high-speed collisions often lead to mass loss, which can unbind the orbit not unlike a supernova \citep[e.g.,][]{Lu+19}. Therefore, we include a stopping condition for orbital energy $\geq 0$ and eccentricity $\geq 1$. We do not consider stars that become unbound after losing mass in a tidal disruption, as proposed by \citet{Manukian+13}.

\section{Numerical Results without Relaxation} \label{sec:results}

\begin{figure*}
	\includegraphics[width=0.8\textwidth]{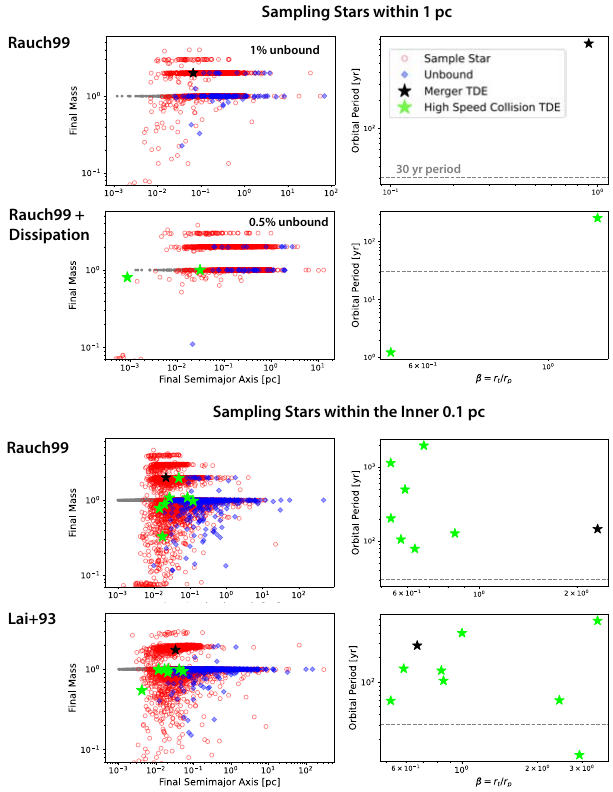}
	\caption{\textbf{Left Column:} We show the final masses of our sample stars versus their final distances from the SMBH for four select simulations without relaxation, allowing us to isolate the effect of collisions. The final mass and distance are either the stars properties when it evolved off the main-sequence or at the end of the $10$~Gyr integration time, whichever is shorter. The grey points are the initial conditions, while red circles are the final sample stars that remained in the cluster. Blue diamonds denote stars which escaped from the cluster, and black and lime stars represent TDEs from different types of collisions, mergers and high-speed collision deflections, respectively. 
    In the upper left hand corner, we indicate the percentage of stars that became unbound orbits. Since unbound orbits have negative semimajor axis, here we plot the absolute value for their final orbits. \textbf{Right Column:} We show the orbital period of the stars when they become TDEs versus $\beta$, the ratio of the tidal radius to the orbit's periapsis. The grey dashed horizontal line marks $30$~yr, approximately the threshold below which we could conveivably observe multiple passages. \textbf{Rows:} Each row corresponds to different initial conditions or collision treatment. \textbf{First row} uses 
    the Rauch99 prescription for mergers and mass loss \textbf{Second row} is identical to the above except we always reduce the post-collision speed in the center of mass frame by $10 \%$ for high-speed collisions. \textbf{The third and fourth rows} use Rauch99 and Lai+93 fitting formulae, respectively. These are higher resolution `zoom-in' simulations where the stars have been sampled on a cusp from $0.001$ to $0.1$~pc. In this radius range, the collision timescale becomes comparable to the lifetime of a solar mass star and increasing our resolution in this range allows us to follow rarer events, such as merger TDEs.}
     \label{fig:norlx}
\end{figure*}

We begin by presenting simulated results without our relaxation prescription in Figure~\ref{fig:norlx}. We run large sets of $10000$ sample stars and let them evolve over $10$~billion years. These simulations allows us to isolate the orbital effects of collisions unobscured by relaxation.
The first two rows of Figure~\ref{fig:norlx} both use the Rauch99 prescription for mass loss and merger conditions. The simulation in the second row, however, tests the potential effects of dissipation during high speed collisions: we decrease the speeds in the stars' center of mass frame by $10 \%$ following each high speed collision.

As described in our methodology section, our sample stars lie between $0.001$ and $1$~pc and share the same spatial and orbital property distributions as the background cluster. The majority of stars reside near $\sim 1$~pc, the edge of the sphere of influence, as expected for a steep density cusp. Only about $500$ stars in our sample, or $5 \%$, lie within $0.1$~pc, where collisions are most common (see Figure~\ref{fig:timescales}), and only about $30$ stars lie within $0.01$~pc, where the collision timescale is orders of magnitude shorter than the relaxation timescale. Sampling our stars on a cusp out to $1$~pc gives a more complete picture of the relative rates of each outcome. If $10$ out of $10000$ stars become gravitationally unbound from the SMBH, for example, we would expect $\sim 0.1\%$ of the total cluster stars within the sphere of influence to share that fate. However, the low number of sample stars in the inner $0.1$ pc also means we do not always resolve unusual collision outcomes in this region, particularly ones that affect $<1$ out of every $500$ stars. In the actual nuclear star cluster, $\sim 10^5$ stars reside within the inner $0.1$ pc. To highlight this region, we include two simulations in the lower half of Figure~\ref{fig:norlx} that follow $10000$ stars in the inner $0.1$~pc of the cluster. These simulations use fitting formulae from Rauch99 and Lai+93, respectively.


The left column of Figure~\ref{fig:norlx} shows the final masses and semimajor axes of the sample stars. Red open circles represent stars at the end of the simulation or their main-sequence lifetime, whichever came first. Grey points represent the initial conditions. Blue diamonds indicate stars that were placed on unbound orbits by high speed collisions. Note that unbound orbits have negative semimajor axes, so we plot the absolute value. We provide percentage of our stars that escape on unbound orbits in the upper right corner. In addition to these unbound orbits, collisions can also place stars on orbits with semimajor axes outside the inner pc.

TDEs from direct collisions are represented by star shaped symbols. The color indicates the type of collision responsible for the radial orbit: low-speed collisions, which we refer to as merger TDEs, are shown in black, while lime represents high speed collisions. 
The right column shows the TDE properties for each simulation: the orbital period of the final orbit on the y-axis and the parameter $\beta$ on the x-axis. This orbit carries the stars into the tidal radius, while $\beta$ quantifies how deeply the orbit penetrates the tidal radius. For stars with mass $< 1 M_\odot$, $0.5< \beta < 1$ indicate a partial disruption, while for masses $> 1 M_\odot$, $0.5< \beta < 2$ indicate a partial disruption\footnote{These choices for full versus partial disruptions are based on simulations in \citep[e.g.][]{Guillochon+13, gafton_tidal_2019, law-smith_stellar_2020, ryu_tidal_2020}}. To guide the eye, grey dashed line marks where the orbital period equals $30$~yr. Partial TDEs with periods less than this threshold could conceivably be observed as a repeating TDE in the next 10 years \citep[given that the first TDE candidates were discovered in the mid 2000s, e.g.][]{gezari_luminous_2009, esquej_candidate_2007}). We discuss the implications for TDEs below.

\subsection{Tidal Disruption Events} \label{sec:TDEs}
We find that stellar collisions can place stars onto nearly radial orbits and produce TDEs. As shown in Figure~\ref{fig:norlx}, these collision TDEs come from the inner $0.1$ pc region, where collisions become important to the evolution of the cluster. This channel can also produce repeating TDEs -- for example the short period TDE in the second row of Figure~\ref{fig:norlx}. 

In order to understand the statistical fluctuations from our fiducial model out to $1$~pc, we ran twenty total simulations using the Rauch99 prescription depicted in the first row of Figure~\ref{fig:norlx}. Of the twenty, twelve produced zero TDEs, seven produced one TDE, and one produced three TDEs, averaging out to roughly $1$ collision TDE per $20,000$ stars. Four of the collision TDEs were from mergers, while the remaining six came from high-speed collisions. We also ran a smaller number of our other simulations and determined that the amount of statistical variation found in our fiducial runs appeared representative across runs with different collision and merger prescriptions. From four runs of the Rauch99 plus dissipation simulation (second row of Figure~\ref{fig:norlx}), one resulted in zero TDEs, one in one TDE, and two in two TDEs. Four of the five TDEs were from high-speed collisions. Similarly, in the third row of the Figure (a zoom-in version of the Rauch99 simulations), high-speed collision TDEs outnumber merger TDEs seven to one. 
Taken altogether, our results indicate that the majority of collision TDEs come from high speed collisions. Merger TDEs occur when two stars with anti-aligned angular orbital angular momenta collide, giving a low angular momentum to the final orbit of the merger product. Part of their rarity comes from the fact that the star must find another star with roughly opposite angular momentum. Another contributing factor is that collisions are most likely to occur near periapsis (see Section~\ref{sec:collisionorbits}), where speeds tend to be higher, making mergers a less likely outcome.

We found even fewer merger TDEs in Lai+93 simulations. Over five runs similar to the first row of Figure~\ref{fig:norlx} but using Lai+93 collision treatments, one had zero TDEs, two had two TDEs, one had three TDEs, and one had four TDEs. None of these eleven TDEs were from mergers. The sole merger from a Lai+93 simulation occurred in the last row ``zoom-in'' simulation in Figure~\ref{fig:norlx}. In previous work, Rauch99 simulations led to more mergers than Lai+93 \citep[see comparison in][]{Rose+23}, though both prescriptions can still produce more massive merger products. The former recipe relies on heuristic arguments, similar to the original study, to determine if a merger occurs, while \citet{Lai+93} include a fitting formula for the merger capture radius.
The relative dearth of merger TDEs in Lai+93 simulations may owe to the fact that this type of collision outcome is generally less common.
Collision-induced mergers are still viable as a mechanism to create TDEs, but they occur at a lower rate. We also note that shallower density profiles will lead to a lower collision rate overall because the timescale is longer (see Figure~\ref{fig:timescales}), in turn reducing the rate of collision TDEs.

Interestingly, for the high speed collision TDEs shown in Figure~\ref{fig:norlx}, the collisions that placed the star on the nearly radial orbit did not result in high mass loss. Stars stripped in collisions do get disrupted, however these stars were often stripped by earlier collisions.  
For example, the $0.78$ and $0.33$~M$_\odot$ in the third row became TDEs following collisions with impact parameters of $0.63 r_c$ and $0.94 r_c$, respectively, that drove off $< 0.1\%$ of the star's mass. TDE-producing collisions tend to result in very low mass loss because it is easier to deflect slower moving stars, but higher speeds are needed for substantial stripping. For similar reasons, most of the high-speed collisions that produced TDEs occured closer to apoapsis even though collisions are more likely to occur near periapsis. All of the high speed collision TDEs in the last row of Figure~\ref{fig:norlx} had initial eccentricities above $0.7$ and most above $0.9$, yet six out of eight of the collisions occurred near apoapsis. Stars move slower at apoapse (relative to at periapse) and are therefore easier to deflect into the the black hole's loss cone (the perpendicular component of $\Delta v \propto 1/v_c$). Lastly, we note a high-speed collision TDE in the third row experienced a merger earlier in its evolution, highlighting the complex collision histories that these stars can have.

\begin{figure}
	\includegraphics[width=0.83\columnwidth]{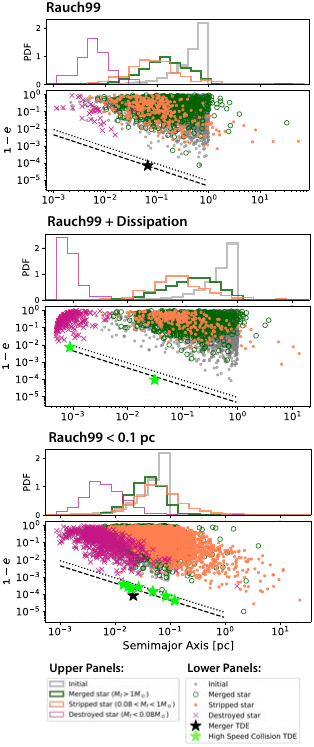}
	\caption{This figure shows semimajor axis versus $1-e$, a quantity proportional to the periapsis distance, of the sample stars from the first three rows of Figure~\ref{fig:norlx}. Systems with $e_f>1$
    (unbound orbits) are ommitted, as are un-collided stars, whose properties cannot change without two-body relaxation.  In the bottom panel of each plot, grey dots represent the initial population. Stars with $M > 1$~M$_\odot$, represented by green circles, must have experienced a merger. Orange dots are stripped stars ($0.08 < M < 1$~M$_\odot$), indicative of a high speed collision. Destroyed stars ($M < 0.08$~M$_\odot$) are marked with pink crosses. TDEs are marked using the same conventions as Figure~\ref{fig:norlx} and lie on nearly radial orbits with extreme eccentricity values. To guide the eye, we plot the full (partial) disruption radii for a $1$~M$_\odot$ star in dashed (dotted) black lines. The upper panel of each plot shows a histogram of the spatial distribution of these subpopulations in the same color scheme as the scatter plot. 
    }
     \label{fig:orbitalproperties_new}
\end{figure}

\subsection{Orbital Changes from Collisions} \label{sec:orbitalshaping}

We consider general trends in the effects of collisions on the orbital properties of the stars. We do this by examining sub-populations of the sample stars based on their final mass. We stress that this classification may not give the full picture of a star's collision history. While all stars with $M_{final}>1$~M$_\odot$ must have undergone a merger -- recall that all stars in \textbf{the} cluster are initially $1$~M$_\odot$ -- and stars with $M_{final}<1$~M$_\odot$ must have lost mass in a high speed collision, stars can experience both types of collisions over their lifetimes. However, final mass still represents the best ``observable'' probe of a star's collision history. We compare the semimajor axis versus $1-e$ of the stars in Figure~\ref{fig:orbitalproperties_new}, the latter being proportional to the periapsis. Grey dots show the initial conditions for all the stars. Green circles represent stars with $M_{final}>1$~M$_\odot$, while orange dots represent stars with $0.08< M_{final}<1$~M$_\odot$. Pink crosses represent stars that experienced enough destructive collisions that their final masses were less than $0.08$~M$_\odot$, in some cases completely destroyed. TDEs and unbound stars are represented using the same symbols as in previous figures. We confirm that the TDEs in our simulation do indeed come from nearly radial orbits. 

Mergers and high speed collisions differ in terms of their orbital outcomes. In the low mass-loss, sticky sphere regime, the orbital angular momentum is conserved, in the isotropic cluster, the angular momenta of the colliding stars can be both aligned or anti-aligned. As a result, merged star orbits can be both more and less eccentric. Conservation of energy demands that mergers always shrink the semimajor axis, while high speed collisions can place stars on both wider or smaller orbits by facilitating an energy exchange between the two stars. $85 \%$ of merged stars in the upper panel of Figure~\ref{fig:orbitalproperties_new} had smaller final semimajor axes compared to their initial orbits. As can be seen in both panels of the figure, there is substantial overlap between the merged and stripped stars, though generally merger products lie at slightly larger radii, where these collisions are more likely to occur.
The few merged stars outside our initial cluster were deflected onto those wide orbits by high-speed collisions 

A roughly equal number of stripped stars moved to more (less) tightly bound orbits. We count unbound stars in the less tightly bound category because their final orbital energies are larger than the initial values. Additionally, $\sim 10$ out of the $10,000$ stars are placed on orbits with semimajor axes outside of $1$~pc. This result suggests that there may be collision-affected stars masquerading outside of the sphere of influence. Including a treatment for dissipation during a high speed collision presents a slightly different story. The orbits of stripped stars are more likely to shrink. Compared to the first row, about ten times as many stars get destroyed ($M_{final} < 0.08$~M$_\odot$) after sinking towards the SMBH, where speeds are higher and collisions more destructive.\footnote{We do not allow collisions to occur within $0.001$ pc, the periapsis of the closest known star to our SMBH \citep[e.g.,][]{Gillessen+17}. The innermost pink crosses in the second row of Figure~\ref{fig:orbitalproperties_new} trace out a curve because the closest possible final orbit of a destroyed star must have its apoapsis near $0.001$ pc.}
We note that in actuality, the orbital properties of stars can be modified by resonant and non-resonant relaxation processes, not accounted for in these results \citep[e.g.,][]{RauchTremaine96,Hopman+06seg,KocsisTremaine11}.


\subsection{Unbound Stars} \label{sec:unbound}

\begin{figure}
	\includegraphics[width=0.99\columnwidth]{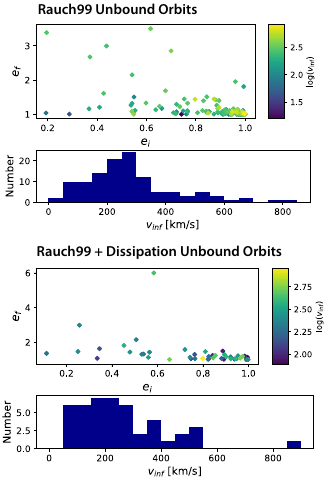}
	\caption{We plot the final versus initial eccentricities for the stars on unbound orbits for the first and second row in Figure~\ref{fig:norlx}. Both the initial and final eccentricities are clustered near $1$, the initial eccentricities slightly under this value and the final eccentricities above. The diamonds are colorcoded by the speed at infinity of the unbound stars, whose distributions are shown in the histograms. We note that this calculation does not take into account any additional contribution to the potential, such as the stellar cluster, beyond the mass of the SMBH. Typical $v_{inf}$ are order $100$~km/s.}
     \label{fig:unboundproperties}
\end{figure}

High speed collisions can unbind stars from the SMBH by transferring energy from one star to another, moving one to a more tightly bound orbit and boosting the other's orbital energy. 
We show properties of the unbound stars in Figure~\ref{fig:unboundproperties} from the simulations in the first and second row of Figure~\ref{fig:norlx}. The initial orbits tend to be eccentric. We also confirmed by inspection of specific cases that the collisions tend to occur near periapsis. For the most part, the final orbits have eccentricity just above unity. We calculate the speed at infinity, $v_{inf}$, for these stars based on their final orbital energy and color code the points in Figure~\ref{fig:unboundproperties} based on its value. Typical speeds range from $\sim 100$ to $600$~km/s, but occasionally one star will have $v_{inf}$ above $1000$~km/s (see Figure~\ref{fig:rlx} for an example). 

Dissipation during the collision can reduce the number of unbound stars.
The second row simulation, which includes some form of dissipation during high speed collisions, resulted in fewer unbound stars overall, though their range of speeds was similar. Specifically, reducing the speed by $10 \%$ post-collision in the frame of the center of mass of the two stars reduced the fraction of unbound stars by about half. We also tested a $50 \%$ reduction in speed for all high speed stellar collisions as described in Section~\ref{sec:HSorbitcalc}. This simulation was a simple way to test the effect of strong dissipation during during the highest speed impacts, yet a population of unbound stars persisted: about $30$ out of $10,000$ stars were ejected, just around $0.3 \%$ (see the middle panel of Figure~\ref{fig:moresimulations} in the Appendix). 

The maximum speeds at infinity suggest that high speed collisions may represent another mechanism to launch hypervelocity stars from galactic nuclei. The most famous of these mechanisms is the Hills Mechanism, in which a binary is disrupted by the SMBH such that one star is ejected at high speed while the other is retained on a tightly bound orbit \citep[e.g.,][]{Hills88,Generozov+20}, though other mechanisms exist \citep[e.g.,][]{YuTremaine,Perets09}. Close encounters between single stars are known to eject stars from dense stellar clusters \citep{Henon69,LinTremaine80}, and in fact \citet{YuTremaine} consider such interactions as a means of producing hypervelocity stars. Omitting collisions, they find a low ejection rate at speeds $\gtrsim 300$~km/s. We consider collisions exclusively and treat them as modified close encounters. While our largest speed is consistent with those generated by the Hill's Mechanism \citep[$\gtrsim 1000$~km/s,][]{Hills88}, about $30 \%$ have $v_{inf} > 300$~km/s. Observations of hypervelocity stars with origins pointing to the Galactic center exhibit a range of speeds, from hundreds to thousands of km/s \citep[e.g.,][to quote from the latter, stars need ejection speed $\gtrsim 750$~km/s from the Galactic center to have $275$~km/s at $20$~kpc]{Brown+05,Brown+18,Koposov+20,GenerozovPerets22}. The $v_{inf}$ of our stars shown in Figure~\ref{fig:unboundproperties} do not account for any additional terms in the potential beyond the SMBH, and we reserve a detailed comparison of our results to observations for future work.

The mass loss fitting formulae from \citet{Rauch99} give very low mass loss for larger impact parameter collisions ($b \gtrsim 0.6 r_c$) at moderately high speeds ($\sim 1000$~km/s); generally, the fractional mass loss in these cases is less than a percent, and several unbound stars in Figure~\ref{fig:norlx} appear as if they have not lost any mass at all. This finding is consistent with Section~\ref{sec:TDEs}: the interactions with the strongest deflections, and therefore orbital changes, tend to have $b \sim r_c$ and therefore result in little mass loss.
We reserve a detailed study of the unbound stars' mass distribution for future work, but some of them may look like stripped stars \citep{Gibson+24}, while many others may simply look like typical stars of their mass and age.

As can be seen in Figure~\ref{fig:norlx}, a maximum of $\sim 1 \%$ of the stars in the inner parsec region may become unbound from the SMBH. This number far exceeds those that become TDEs through high speed collisions. Furthermore, this outcome represents a surprisingly high fraction of stars that experience such collisions: about a fifth of stars that experience high speed collisions become unbound. 
This high number may owe to a few conspiring conditions:

For reasons discussed in Section~\ref{sec:stoppingconditions} and as supported by Figure~\ref{fig:unboundproperties}, unbound stars tend come from high speed collisions near the periapsis of eccentric orbits. Near periapsis, a small boost in the frame of the SMBH can tip the star's speed over the escape speed.
However, the merger-to-high speed collision boundary also lies around $600$~km/s. The velocity dispersion reaches this value around $0.02$~pc, and the vast majority of stars reside outside this distance. In consequence, most high speed collisions can only occur near periapsis, where speeds are high enough to no longer be in the merger regime and also where unbinding the star from the SMBH becomes a more favorable outcome. Our thermal initial eccentricity distribution ensures that plenty of stars begin on very eccentric orbits (see Appendix~\ref{sec:uniformecc} for the impact of the eccentricity distribution).


\section{Numerical Results with Relaxation} 
\label{sec:withrelaxation}

\begin{figure*}
	\includegraphics[width=0.99\textwidth]{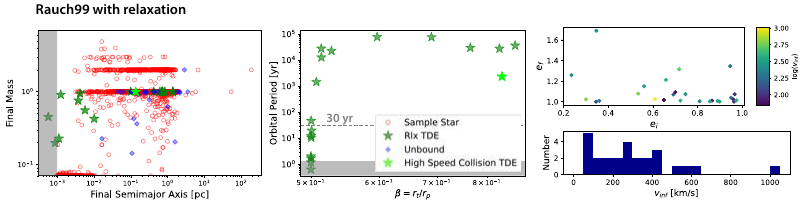}
	\caption{We show a simulation that uses the Rauch99 prescription that includes relaxation. There are $4000$ sample stars in this simulation. About $0.7\%$ became unbound, and one became a TDE immediately post-collision. Some other collision-affected stars also became TDEs; collisions would have affected their orbital evolution, but it was the relaxation prescription that pushed them into the tidal radius. The symbols are consistent with those in our other figures, with the addition of the dark green stars to represent TDEs from our relaxation prescription. Indicated by the grey shaded region, we place a caveat on the dark green stars (relaxation TDEs) with final semimajor axis within $0.001$~pc, also the TDEs with the shortest orbital period. The applicability of our relaxation prescription within this region is uncertain.
    }
     \label{fig:rlx}
\end{figure*}

Now that we have built a physical picture of how collisions shape the stellar cluster, we turn on our prescription for two-body relaxation in the code. These simulations take longer to run, so the sample size of the sample stars presented herein number at $4000$ instead of $10000$. We present results in Figure~\ref{fig:rlx}. The right panel shows the properties of the unbound stars in the style of Figure~\ref{fig:unboundproperties}. About $0.7 \%$ of the stars in this simulation became unbound during a high-speed collision, a fraction that is consistent to order of magnitude with our simulation without relaxation. The escape velocities are also consistent with earlier findings, with one star clearly in the hypervelocity regime. The left and middle panels of Figure~\ref{fig:rlx} parallel Figure~\ref{fig:norlx}, highighting the TDEs. The symbols are the same as in previous figures, with one addition: stars that became TDEs through two body relaxation are in dark green. The orbital evolution of these stars where still affected by collisions, as can be seen by their masses. In fact, roughly half of the TDEs from two-body relaxation were collision-affected, though only $6$ changed their mass by over $10\%$. 
We place a caveat on the TDEs inside $0.001$~pc, the region shaded grey in the Figure. The closest known star to the Milky Way's SMBH has a periapsis just under $0.001$~pc \citep[e.g.,][]{Gillessen+17}. Because the density of stars at these radii is uncertain, the applicability of our two-body relaxation prescription within this radius is also uncertain. If we assume there are stars at these radii that can produce these TDEs, we find that some of these TDEs are partial disruptions with orbital timescales $\lesssim 30$ years, and could be observed as repeating TDEs. 
Consistent with our findings in Section~\ref{sec:TDEs}, the collision that placed a star onto a disrupting orbit did not result in much mass loss (less than a percent). However, even without stripping mass, the collision may still inflate the star's outer layers, potentially affecting the observational signature \citep[e.g.,][]{macleod_spoon-feeding_2013,Gibson+24}. Additionally, we may be missing disruptions from stars that are inflated post-collision as we are not increasing the tidal radius to account for potential radius inflation.

Based on our results in Figure~\ref{fig:rlx} and Figure~\ref{fig:norlx}, we estimate a collision-TDE rate from $10^{-8}$ to $10^{-7}$ per galaxy per year. This is of order $100-1000\times$ lower than the overall TDE rate in the relevant black hole mass range \citep[][]{hannah_counting_2024, chang_rates_2024}. However, if we focus on our final simulation that includes both collisions and relaxation, our rate of TDEs of stars that have at some point {\it experienced} significant stripping ($\gtrsim 30\%$ mass loss) or mergers from collisions is at least a factor of 3 higher  (not including TDEs within 0.001 pc, see Figure~\ref{fig:rlx}), and therefore is $\sim 30 - 300\times$ lower than the overall TDE rate. LSST is expected to observe $\gtrsim 30,000$ TDEs \citep[][]{bricman_prospects_2020}, and therefore even if only 1\% of those events get detailed follow-up, we may still observe one to a few of these collision-induced TDEs with LSST, and even more collision-{\it affected} TDEs. 

Furthermore, other GN with larger ($\sim 5 \times 10^6-10^8$ M$_\odot$) SMBHs can have shorter collision timescales \citep[see, e.g., Figure 1 in][]{FreitagBenz}. The collision rate may therefore be lower in the specific system we looked at, but higher in others with a more massive central SMBH. The overall TDE rate decreases over this mass range both because of dynamical arguments related to the density profile \citep[][]{WangMerritt04} and due to the SMBH's event horizon approaching the tidal radius. However, observational evidence exists for $> 10$ TDEs from black holes $>10^7$ M$_\odot$ and a few from black holes $\gtrsim 10^8$ M$_\odot$ \citep[see e.g.][]{wevers_black_2017, wevers_black_2019, mockler_weighing_2019, van_velzen_optical-ultraviolet_2020, hammerstein_integral_2023-1}. Collisions may impact more stars in these nuclear star clusters, and in turn, the percentage of collision-affected TDEs may increase in this mass range even as the total number of TDEs decreases.
The interplay of relaxation and collisions merit a more thorough examination, and we reserve a detailed study for future work.

\section{Conclusions} \label{sec:conclusions}

Collisions between main-sequence stars occur in the nuclear star cluster and become common within the inner $0.1$~pc \citep[e.g.,][]{RoseMacLeod24}. In this proof-of-concept study, we examine the orbital changes that result from these collisions. Specifically, we assess whether physical collisions can contribute to the production of TDEs and ejected stars. Collisions in galactic nuclei can be understood in two types.

In the first type, the relative speed between the stars is low and the collision results in a merger. Little mass loss occurs \citep[e.g.,][]{Lai+93} and we can treat the stars as sticky spheres \citep[e.g.,][]{Kremer+20,Gonzalez+21}. Conservation of momentum allows us to calculate the final orbit of the merged star. The second type of collision occurs at speeds that exceed the escape velocity from the stars. While the high speeds ensure that the stars remain unbound from each other, they can also drive mass loss from the stars, in some cases producing a stripped star \citep[e.g.,][]{Lai+93,Rauch99,FreitagBenz,Gibson+24}. We use fitting formulae from \citet{Rauch99} and \citet{Lai+93} to calculate the mass loss. 
We are examining these collisions in greater detail using SPH simulations in forthcoming work, which we will use to tune our treatment of the collisions.

In this proof-of-concept study, we use limiting cases and heuristic arguments to determine the final orbits from high speed collisions. If the stars were point masses, these interactions would unfold as a hyperbolic encounter in the center of mass frame of the stars. As an upper limit, we could calculate the deflection angle using Eq.~\ref{eq:theta}. However, while a grazing collision approaches the limit given by Eq.~\ref{eq:theta}, a head-on collision should not result in any deflection at all. We therefore adopt Eq.~\ref{eq:colltheta}, which meets both criteria. We also test the role of dissipation by reducing the speeds of the stars post-collision in their center of mass frame.

Our simulations follow a sample of sample stars embedded in a fixed, uniform cluster. We test a variety of cluster conditions
and find the following:
\begin{enumerate}
    \item \textbf{Orbital Effects of Collisions:} We find that collision-induced mergers work to shrink the orbits, while high-speed collisions can facilitate an energy exchange between stars, allowing them to move to both wider and smaller orbits. There is overlap in the populations that experience these kinds of collisions. Both stripped stars and merger products may be scattered to a much wider orbit, with semimajor axis even outside of the SMBH's sphere of influence, by a later high speed collision. If high-speed collisions are found to be significantly dissipative, however, affected stars also sink to smaller orbits about the SMBH.

    \item \textbf{Tidal disruption events:} We find that both high and low speed collisions can produce TDEs by placing stars on highly eccentric orbits. 
    Our results show that collision-affected stars that are stripped or have experienced mergers can be delivered to the tidal radius both by the same dynamical process, and by standard two-body relaxation that affects the population at large. We predict the rates of TDEs from collisionally-affected stars to be within a factor of $\sim 30-300$ the overall TDE rate from two-body relaxation (see Section~\ref{sec:withrelaxation}), and therefore some of these events should be observed by LSST \citep[e.g.][]{bricman_prospects_2020}.
    The TDEs immediately post-collision versus after the star has relaxed may look quite different \citep[see discussion of imminent versus eventual TDEs in][]{Gibson+24}, adding to the richness of the observations that this dynamical process can yield.

    
    \item \textbf{Unbound stars:} High speed collisions can place stars on unbound orbits, especially if the collisions occur near the periapsis of an eccentric orbit. As evidenced by the low mass loss in Figure~\ref{fig:norlx}, these collisions tend to have larger impact parameter. 
    The unbound stars persist even if the collisions are dissipative, albeit in lower numbers. We reserve a precise examination of their rates and properties for future work, though some may look like stripped stars \citep{Gibson+24}.
    In the optimistic case, stellar collisions represent a mechanism to launch hypervelocity stars from galactic nuclei, joining the list of existing mechanisms \citep[e.g.,][]{Hills88,YuTremaine,GinsburgLoeb,Perets09,Perets09a,Generozov+20}. The Hills Mechanism elegantly explains both the origins of the S-star cluster, young-seeming massive stars in the vicinity of the SMBH, and hypervelocity stars using the same dynamical process \citep[e.g.,][]{Hills88,Ghez+03,GinsburgLoeb,Perets+07,Madigan+09,Lockman+09,Generozov+20,Generozov21}. Collisions present another possibility, where S-stars are the high-mass tail of the merger products created by successive low-speed collisions \citep[see e.g.,][]{Rose+23}, and hypervelocity stars trace to high speed collisions near periapsis.
    However, the viability of this ejection mechanism still faces tests in the form of dissipation during collisions and a stellar mass function.
\end{enumerate}

\begin{acknowledgments}
We thank the anonymous referee for their feedback, which helped strengthen this study. We thank Fred Rasio, Enrico Ramirez-Ruiz, Jamie Lombardi, Fulya Kiroglu, Charles Gibson, Claire Ye, Christopher O'Connor, and Jiaru Li for invaluable discussion and input. SR thanks the CIERA Lindheimer Fellowship for support. BM thanks the Carnegie CTAC postdoctoral fellowship for support. This project began while SCR and BM were at the Aspen Center for Physics, which is supported by NSF grant PHY-2210452. This research was supported in part through the computational resources and staff contributions provided for the Quest high performance computing facility at Northwestern University which is jointly supported by the Office of the Provost, the Office for Research, and Northwestern University Information Technology.

\end{acknowledgments}




\appendix

\section{Additional Simulations} \label{sec:supplementalfigures}

Here we include additional simulations.
First, we show an additional Rauch99 simulation, similar to the first row of Figure~\ref{fig:norlx}, that exhibits both types of collision TDEs and an escaped star with $v_{inf}>1000$~km/s. Of the Rauch99 no relaxation simulations, this one produced the most TDEs. Second, we plot a run where we reduce the post-collision speed of all high speed collisions by 50\% to test the post-collision orbital evolution of the most extremely dissipative collisions. The vast majority of high speed collisions will not be this dissipative, however head on collisions with small impact parameters could result in very high dissipation of kinetic energy. We find, perhaps unsurprisingly, that after experiencing these collisions, the stars sink into the SMBHs potential well and are disrupted at very high rates. This further motivates hydro studies calculating the dissipation per collision for high velocity, small impact parameter collisions, because despite being rare, they could contribute an outsized amount to the  number of collision-induced TDEs. Lastly, we show the results of a version of our fiducial run (without relaxation) with the Lai+93 collision treatment (a plot of the zoom-in simulation of the high collision probability is included in Figure~\ref{fig:norlx} of the paper).


\begin{figure*}
	\includegraphics[width=0.9\textwidth]{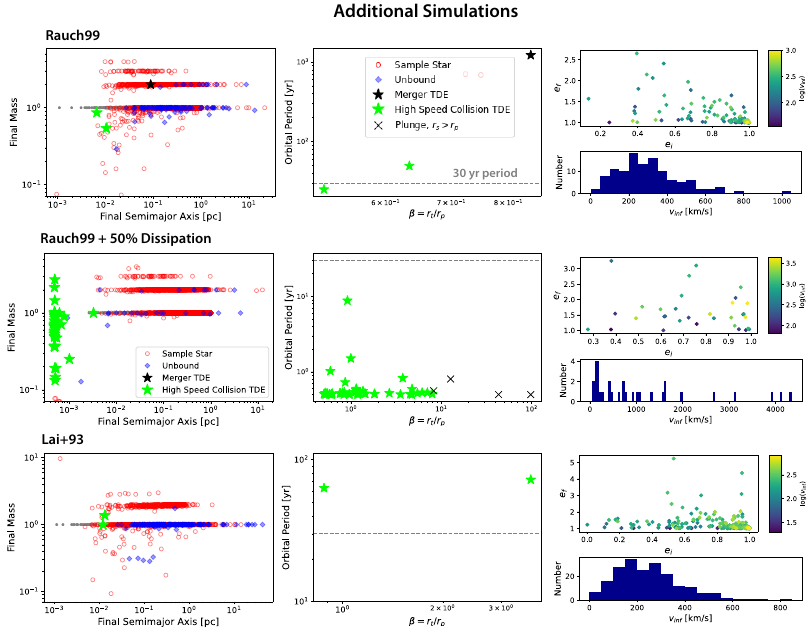}
	\caption{This figure has the same form as Figure~\ref{fig:rlx}. The symbols are the same as in previous figures, except with the addition of crosses, which represent stars placed on such radial orbits that they were deposited within the schwarzschild radius of the SMBH. In the first row, we show another example of a Rauch99 simulation similar to the one in the first row of Figure~\ref{fig:norlx}. In the second row, we show a Rauch99 with dissipation during high-speed collisions: we scaled the post-collision speed by $50 \%$ in center of mass frame following high-speed, destructive collision cases. The last row shows a Lai+93 simulation for a sample of stars out to $1$~pc. }
     \label{fig:moresimulations}
\end{figure*}

\section{Role of the Initial Eccentricity Distribution} \label{sec:uniformecc}

Less eccentric orbits on average should result in fewer unbound stars. We test this hypothesis by drawing the eccentricities of sample stars from a uniform distribution (average $e = 0.5$) instead of a thermal one (average $e = 0.67$). 
We find that the number of unbound stars is less by about a factor of $2$ compared to simulations with a thermal initial eccentricity distribution. These results come from reducing the number of stars with speeds at periapsis, where collisions are often in the high speed collision regime.

\bibliography{sample631}{}
\bibliographystyle{aasjournal}



\end{document}